# A Cell Dynamical System Model for Simulation of Continuum Dynamics of Turbulent Fluid Flows

A. M. Selvam and S. Fadnavis

Email: amselvam@gmail.com

Website: http://www.geocities.com/amselvam

## 1. INTRODUCTION

Atmospheric flows exhibit long-range spatiotemporal correlations manifested as the *fractal* geometry to the global cloud cover pattern concomitant with inverse power-law form for power spectra of temporal fluctuations of all scales ranging from turbulence (millimeters-seconds) to climate (thousands of kilometers-years) (Tessier et. al., 1996) Long-range spatiotemporal correlations are ubiquitous to dynamical systems in nature and are identified as signatures of *self-organized criticality* (Bak et. al., 1988) Standard models for turbulent fluid flows in meteorological theory cannot explain satisfactorily the observed *multifractal* (space-time) structures in atmospheric flows. Numerical models for simulation and prediction of atmospheric flows are subject to *deterministic chaos* and give unrealistic solutions. *Deterministic chaos* is a direct consequence of round-off error growth in iterative computations. Round-off error of finite precision computations doubles on an average at each step of iterative computations (Mary Selvam, 1993). Round-off error will propagate to the mainstream computation and give unrealistic solutions in numerical weather prediction (NWP) and climate models which incorporate thousands of iterative computations in long-term numerical integration schemes. A recently developed non-deterministic cell dynamical system model for atmospheric flows (Mary Selvam, 1990; Mary Selvam et. al., 1996) predicts the observed self-organized criticality as intrinsic to quantumlike mechanics governing flow dynamics. Further, the fractal space-time structure to the stringlike atmospheric flow trajectory is resolved into a continuum of eddies. The eddy circulations obey *Kepler*'s third law of planetary motion and therefore eddy inertial masses obey *Newton's inverse square law of gravitation* on all scales from microscopic to macroscale (Selvam and Fadnavis, 1998a, b). El Naschie (1997) has discussed the fractal structure to space-time and also states that fractalisation of microspace is the origin of gravity.

## 2. A NON-DETERMINISTIC CELL DYNAMICAL SYSTEM MODEL FOR FLUID FLOWS: A STRING THEORY FOR FRACTAL SPACETIME

Based on Townsend's (1956) concept that large eddies are the envelopes of enclosed turbulent eddy circulations, the relationship between the large and turbulent eddy circulation speeds (*W* and *w₊*) and radii ( *R* and *r* ) respectively is given as

$$W^2 = \frac{2}{\pi}\frac{r}{R}w_*^2 \qquad (1)$$

Since the large eddy is the integrated mean of enclosed turbulent eddy circulations, the eddy energy (kinetic) spectrum follows statistical normal distribution. Therefore, square of the eddy amplitude or the variance represents the probability. Such a result that the additive amplitudes of eddies, when squared, represent the probability densities is obtained for the subatomic dynamics of quantum systems such as the electron or photon (Maddox, 1988). Atmospheric flows, therefore, follow quantumlike mechanical laws. Incidentally, one of the strangest things about physics is that we seem to need two different kinds of mechanics, quantum mechanics for microscopic dynamics of quantum systems and classical mechanics for macroscale phenomena (Rae, 1988). The above visualization of the unified network of atmospheric flows as a quantum system is consistent with Grossing's (Grossing, 1989) concept of quantum systems *as order out of chaos* phenomena. Order and chaos have been reported in strong fields in quantum systems (Brown, 1996).

The square of the eddy amplitude $W^2$ represents the kinetic energy *E* given as (from equation .1)

$$E = H\nu \qquad (2)$$

Where $\nu$ (proportional to *1/R*) is the frequency of the large eddy and *H* is a constant equal to $\frac{2}{\pi}rw_*^2$ for growth of large eddies sustained by constant energy input proportional to $w_*^2$ from fixed primary small scale eddy fluctuations. Energy content of eddies is therefore similar to quantum systems which can possess only discrete quanta or packets of energy content $h\nu$ where *h* is a universal constant of nature (*Planck's constant*) and $\nu$ is the frequency in cycles per second of the electromagnetic radiation. The relative phase angle between large and turbulent eddies is equal to *r/R* and is directly proportional to $W^2$ (equation .1). The phase angle therefore represents variance and also there is progressive increase in phase with increase in wavelength. The above relationship between phase angle, variance and frequency has been identified as *Berry's Phase* (Berry, 1988) in the subatomic dynamics of quantum systems. Berry's phase has been identified in atmospheric flows (Mary Selvam et. al., 1996; Mary Selvam, 1997; Mary Selvam and Suvarna Fadnavis, 1998a).

Writing equation (1) in terms of the periodicities *T* and *t* of large and small eddies respectively, where

$$T = \frac{2\pi R}{W}$$

and

$$t = \frac{2\pi r}{w_*}$$

We obtain

$$\frac{R^3}{T^2} = \frac{2}{\pi}\frac{r^3}{t^2} \tag{3}$$

Equation (3) is analogous to *Kepler*'s third law of planetary motion, namely, the square of the planet's year (period) to the cube of the planet's mean distance from the *Sun* is the same for all planets (Weinberg,1993). Newton developed the idea of an inverse square law for gravitation in order to explain Kepler's laws, in particular, the third law. Kepler's laws were formulated on the basis of observational data and therefore are of empirical nature. A basic physical theory for the inverse square law of gravitation applicable to all objects, from macroscale astronomical objects to microscopic scale quantum systems is still lacking. The model concepts are analogous to a string theory (Kaku,1997) where, superposition of different modes of vibration in stringlike energy flow patterns result in material phenomena with intrinsic quantumlike mechanical laws which incorporate inverse square law for inertial forces, the equivalent of gravitational forces, on all scales of eddy fluctuations from macro- to microscopic scales. The cumulative sum of centripetal forces in a hierarchy of vortex circulations may result in the observed inverse square law form for gravitational attraction between inertial masses (of the eddies).

**WAVE-PARTICLE DUALITY**
wave-trains in atmospheric flows and cloud formation

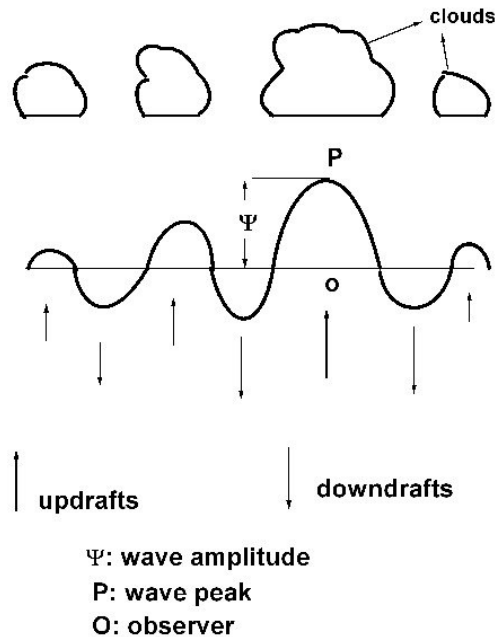

$\Psi$: wave amplitude
P: wave peak
O: observer

Figure 1. Wave-particle duality in atmospheric flows.

The apparent paradox of wave-particle duality in microscopic scale quantum systems (Rae,1988) is however physically consistent in the context of macroscale atmospheric flows since the bi-directional energy flow structure of a complete atmospheric eddy results in the formation of clouds in updraft regions and dissipation of clouds in downdraft regions. The commonplace occurrence of clouds in a row is a manifestation of *wave-particle duality* in the macroscale quantum system of atmospheric flows (Fig. 1).

The above-described analogy of quantumlike mechanics for atmospheric flows is similar to the concept of a subquantum level of fluctuations whose space-time organization gives rise to the observed manifestation of subatomic phenomena, i.e. quantum systems as order out of chaos phenomena.

Puthoff (1989) has also put forth the concept of "gravity as a zero-point fluctuation force". The vacuum zero-point fluctuation (electromagnetic) energy manifested in the *Casimir effect* is analogous to the turbulent scale fluctuations whose spatial integration results in coherent large eddy structures. El Naschie has proposed in a series of papers (1997) that *Cantorian-fractal* conception of spacetime may effect reconciliation between quantum mechanics and gravity.

*2.1. Model Predictions*

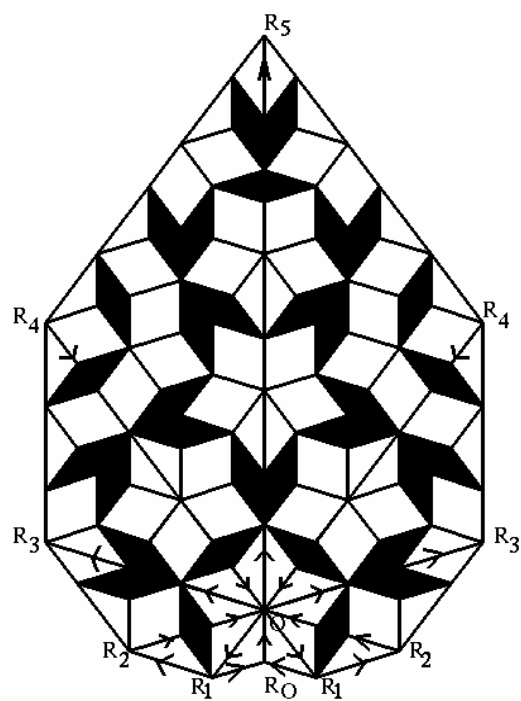

Figure 2. The quasiperiodic *Penrose* tiling pattern.

(a) Atmospheric flows trace an overall logarithmic spiral trajectory $OR_0R_1R_2R_3R_4R_5$ simultaneously in clockwise and anti-clockwise directions with the quasi-periodic *Penrose tiling pattern* (Steinhardt, 1997) for the internal structure Fig. 2.

(b) Conventional continuous periodogram power spectral analyses of such spiral trajectories will reveal a continuum of periodicities with progressive increase in phase.

(c) The broadband power spectrum will have embedded dominant wavebands, the bandwidth increasing with period length. The peak periods $E_n$ in the dominant wavebands will be given by the relation

$$E_n = T_S(2+\tau)\tau^n \quad (4)$$

Where $\tau$ is the *golden mean* equal to $(1+\sqrt{5})/2$ [$\cong 1.618$] and $T_s$ is the primary perturbation time period, for example, the solar powered annual cycle (summer to winter) of solar heating in a study of interannual climate variability. Ghil (1994) reports that the most striking feature in climate variability on all time scales is the presence of sharp peaks superimposed on a continuous background. The model predicted periodicities are *2.2*, *3.6*, *5.8*, *9.5*, *15.3*, *24.8*, *40.1* and *64.9* years for values of *n* ranging from *-1* to *6*. Periodicities close to model predicted have been reported (Burroughs, 1992)

(d) The overall logarithmic spiral flow structure is given by the relation

$$W = \frac{w_*}{k}\ln z \quad (5)$$

Where the constant **k** is the steady state fractional volume dilution of large eddy by inherent turbulent eddy fluctuations. The constant **k** is equal to $1/\tau^2$ ($\cong 0.382$) and is identified as the universal constant for deterministic chaos in fluid flows. The steady state emergence of fractal structures is therefore equal to

$$1/k \cong 2.62 \quad (6)$$

The model predicted logarithmic wind profile relationship such as equation (5) is a long-established (observational) feature of atmospheric flows in the boundary layer, the constant **k**, called the *Von Karman*'s constant has the value equal to *0.38* as determined from observations. Historically, equation (5), basically an empirical law known as the *universal logarithmic law of the wall*, first proposed in the early 1930s by pioneering aerodynamicists *Theodor von Karman* and *Ludwig Prandtl*, describes shear forces exerted by turbulent flows at boundaries such as wings or fan blades or the interior wall of a pipe. *The law of the wall* has been used for decades by engineers in the design of aircraft, pipelines and other structures (Cipra, 1996).

In equation (5), **W** represents the standard deviation of eddy fluctuations, since **W** is computed as the instantaneous r.m.s. (root mean square) eddy perturbation amplitude with reference to the earlier step of eddy growth. For two successive stages of eddy growth starting from primary perturbation $w_*$, the ratio of the standard deviations $W_{n+1}$ and $W_n$ is given from equation (5) as *(n+1)/n*. Denoting by $\sigma$ the standard deviation of eddy fluctuations at the reference level *(n=1)* the standard deviations of eddy fluctuations for successive stages of eddy growth are given as integer multiples of $\sigma$, i.e., $\sigma$, $2\sigma$, $3\sigma$, etc. and correspond respectively to

Statistical normalized standard deviation **t** = *0*,*1*,*2*,*3*, **etc**. (7)

The conventional power spectrum plotted as the variance versus the frequency in log-log scale will now represent the eddy probability density on logarithmic scale versus the standard deviation of the eddy fluctuations on linear scale since the logarithm of the eddy wavelength represents the standard deviation, i.e. the r.m.s. value of eddy fluctuations (5). The r.m.s. value of eddy fluctuations can be represented in terms of statistical normal distribution as follows. A normalized standard deviation *t=0* corresponds to cumulative percentage probability density equal to *50* for the mean value of the distribution. Since the logarithm of the wavelength represents the r.m.s. value of eddy fluctuations the normalized standard deviation **t** is defined for the eddy energy as

$$t = \frac{\log L}{\log T_{50}} - 1 \qquad (8)$$

Where $L$ is the period in years and $T_{50}$ is the period up to which the cumulative percentage contribution to total variance is equal to $50$ and $t = 0$. The variable $\log T_{50}$ also represents the mean value for the r.m.s. eddy fluctuations and is consistent with the concept of the mean level represented by r.m.s. eddy fluctuations. Spectra of time series of meteorological parameters when plotted as cumulative percentage contribution to total variance versus $t$ have been shown to follow the model predicted universal spectrum (Mary Selvam and Suvarna Fadnavis 1998a)

(e) Mary Selvam (1993) has shown that equation (1) represents the universal algorithm for deterministic chaos in dynamical systems and is expressed in terms of the universal *Feigenbaum*'s (Feigenbaum, 1980) *constants* ***a*** and ***d*** as follows. The successive length step growths generating the eddy continuum $OR_0R_1R_2R_3R_4R_5$ analogous to the period doubling route to chaos (growth) is initiated and sustained by the turbulent (fine scale) eddy acceleration $w_*$ which then propagates by the inherent property of inertia of the medium of propagation. Therefore, the statistical parameters *mean*, *variance*, *skewness* and *kurtosis* of the perturbation field in the medium of propagation are given by $w_*$, $w_*^2$, $w_*^3$ and $w_*^4$ respectively. The associated dynamics of the perturbation field can be described by the following parameters. The perturbation speed $w_*$ (motion) per second (unit time) sustained by its inertia represents the mass, $w_*^2$ the acceleration or force, $w_*^3$ the angular momentum or potential energy, and $w_*^4$ the spin angular momentum, since an eddy motion has an inherent curvature to its trajectory.

It is shown that *Feigenbaum's* constant ***a*** is equal to (Mary Selvam, 1993)

$$a = \frac{W_2 R_2}{W_1 R_1} \qquad (9)$$

Where the subscripts $1$ and $2$ refer to two successive stages of eddy growth. *Feigenbaum's* constant ***a*** as defined above represents the steady state emergence of fractional *Euclidean* structures. Considering dynamical eddy growth processes, *Feigenbaum's* constant ***a*** also represents the steady state fractional outward mass dispersion rate and $a^2$ represents the energy flux into the environment generated by the persistent primary perturbation $w_*$. Considering both clockwise and counterclockwise rotations, the total energy flux into the environment is equal to $2a^2$. In statistical terminology, $2a^2$ represents the variance of fractal structures for both clockwise and counterclockwise rotation directions.

The *Feigenbaum's* constant ***d*** is shown to be equal to (Mary Selvam, 1993)

$$d = \frac{W_2^4 R_2^4}{W_1^4 R_1^4} \qquad (10)$$

and represents the fractional volume intermittency of occurrence of fractal structures for each length step growth. *Feigenbaum's* constant ***d*** also represents the relative spin angular momentum of the growing large eddy structures as explained earlier.

Equation (1). may now be written as

$$2\frac{W^2 R^2}{w_*^2 (dR^2)} \equiv \pi \frac{W^4 R^3}{w_*^4 (dR^3)}$$

(11)

Where **dR** equal to **r** represents the incremental growth in radius for each length step growth, i.e. **r** relates to the earlier stage of eddy growth.

Substituting the *Feigenbaum's constants* **a** and **d** defined above (9 and 10) equation (11) can be written as

$$2a^2 = \pi d \qquad (12)$$

Where $\pi d$, the relative volume intermittency of occurrence contributes to the total variance $2a^2$ of fractal structures.

In terms of eddy dynamics, the above equation states that during each length step growth, the energy flux into the environment equal to $2a^2$ contributes to generate relative spin angular momentum equal to $\pi d$ of the growing fractal structures.

It was shown at equation (6) above that the steady state emergence of fractal structures in fluid flows is equal to **1/k** (= $\tau^2$) and therefore the *Feigenbaum's constant* **a** is equal to

$$a = \tau^2 = 1/k = 2.62 \qquad (13)$$

(f) The relationship between *Feigenbaum's constant* **a** and statistical normal distribution for power spectra is derived in the following.

The steady state emergence of fractal structures is equal to the *Feigenbaum's constant* **a** (equation 6). The relative variance of fractal structure for each length step growth is then equal to $a^2$. The normalized variance $1/a^{2n}$ will now represent the statistical normal probability density for the $n^{th}$ step growth according to model predicted quantumlike mechanics for fluid flows. Model predicted probability density values **P** are computed as

$$P = \tau^{-4n} \qquad (14)$$

or

$$P = \tau^{-4t} \qquad (15)$$

where **t** is the normalized standard deviation (equation 7) and are in agreement with statistical normal distribution as shown in Table 1 below.

Table 1: Model predicted and statistical normal probability density distributions

| growth step | normalized std dev | probability densities | |
|---|---|---|---|
| n | t | model predicted $P = \tau^{-4t}$ | statistical normal distribution |
| 1 | 1 | .1459 | .1587 |
| 2 | 2 | .0213 | .0228 |
| 3 | 3 | .0031 | .0013 |

(g) The power spectra of fluctuations in fluid flows can now be quantified in terms of universal *Feigenbaum's constant* **a** as follows.

The normalized variance and therefore the statistical normal distribution is represented by (from equation 14)

$$P = a^{-2t} \tag{16}$$

Where $P$ is the probability density corresponding to normalized standard deviation $t$. The graph of $P$ versus $t$ will represent the power spectrum. The slope $S$ of the power spectrum is equal to

$$S = \frac{dP}{dt} \approx -P \tag{17}$$

The power spectrum therefore follows inverse power law form, the slope decreasing with increase in $t$. Increase in $t$ corresponds to large eddies (low frequencies) and is consistent with observed decrease in slope at low frequencies in dynamical systems.

(h) The fractal dimension $D$ can be expressed as a function of the universal *Feigenbaum's constant a* as follows.

The steady state emergence of fractal structures is equal to $a$ for each length step growth (7 & 13) and therefore the fractal structure domain is equal to $a^m$ at $m^{th}$ growth step starting from unit perturbation. Starting from unit perturbation, the fractal object occupies spatial (two dimensional) domain $a^m$ associated with radial extent $\tau^m$ since successive radii follow *Fibonacci* number series. The fractal dimension $D$ is defined as

$$D = \frac{d \ln M}{d \ln R}$$

where $M$ is the mass contained within a distance $R$ from a point in the fractal object. Considering growth from $n^{th}$ to $(n+m)^{th}$ step

$$d \ln M = \frac{dM}{M} = \frac{a^{n+m} - a^n}{a^n} = a^m - 1 \tag{18}$$

Similarly

$$d \ln R = \frac{dR}{R} = \frac{\tau^{n+m} - \tau^n}{\tau^n} = \tau^m - 1 \tag{19}$$

Therefore the fractal dimension $D$ is given as

$$D = \frac{\tau^{2m} - 1}{\tau^m - 1} = \tau^m + 1 \tag{20}$$

The fractal dimension increases with the number of growth steps. The dominant wavebands increase in length with successive growth steps. The fractal dimension $D$ indicates the number of periodicities which superimpose to give the observed four dimensional space-time structure to the flow pattern. The above concept of dimension for real world spacetime patterns is consistent with El Naschie's (1997) interpretation of dimensions for superstring theories in particle physics, namely a string rotates in ordinary space and only uses the extra dimensions for vibrations which simulate particle masses.

(i) The relationship between *fine structure constant*, i.e. the eddy energy ratio between successive dominant eddies and Feigenbaum's constant $a$ is derived as follows.

$2a^2$ = relative variance of fractal structure (both clockwise and anticlockwise rotation) for each growth step.

For one dominant large eddy (Fig. 2) $OR_0R_1R_2R_3R_4R_5$ comprising of five growth steps each for clockwise and counterclockwise rotation, the total variance is equal to

$$2a^2 \times 10 = 137.07 \tag{21}$$

For each complete cycle (comprising of five growth steps each) in simultaneous clockwise and counterclockwise rotations, the relative energy increase is equal to *137.07* and represents the *fine structure constant* for eddy energy structure.

Incidentally, the *fine structure constant* in atomic physics (Omnes, 1994), designated as $\alpha^{-1}$, a dimensionless number equal to *137.03604*, is very close to that derived above for atmospheric eddy energy structure. This fundamental constant has attracted much attention and it is felt that quantum mechanics cannot be interpreted properly until such time as we can derive this physical constant from a more basic theory.

(j) The ratio of proton mass **M** to electron mass $m_e$, i.e., $M/m_e$ is another fundamental dimensionless number which also awaits derivation from a physically consistent theory. $M/m_e$ determined by observation is equal to about *2000*. In the following it is shown that ratio of energy content of large to small eddies for specific length scale ratios is equivalent to $M/m_e$.

From Equation (21),

The energy ratio for two successive dominant eddy growth = $(2a^2 \times 10)^2$

Since each large eddy consists of five growth steps each for clockwise and anticlockwise rotation,

The relative energy content of large eddy with respect to primary circulation structure inside this large eddy

$$= (2a^2 \times 10)^2/10$$
$$\cong 1879$$

The cell dynamical system model concepts therefore enable physically consistent derivation of fundamental constants which define the basic structure of quantum systems. These two fundamental constants could not be derived so far from a basic theory in traditional quantum mechanics for subatomic dynamics (Omnes, 1994).

### 4. CONCLUSION

The cell dynamical system model presented in the paper is basically a string theory applicable to all dynamical systems ranging from macroscale atmospheric flows to subatomic scale quantum systems. The four dimensional real world spacetime continuum fluctuations are manifestation of the superimposition of a hierarchical continuum of eddy circulations, whose centripetal accelerations add cumulatively to represent the inertial mass, which is equivalent to gravitational mass.

*Acknowledgements*

**REFERENCES**

Bak, P. C, C. Tang and K. Wiesenfeld, Self-organized criticality, *Phys. Rev.* **A 38**, 364-374 (1988).

Berry, M. V., The geometric phase, *Sci. Amer.* **Dec**., 26-32 (1988).

Brown, J., Where two worlds meet, *New Scientist*, **18 May**, 26-30 (1996).

Burroughs, W. J., *Weather Cycles: Real or Imaginary?*, Cambridge University Press, Cambridge (1992).

Cipra, B., A new theory of turbulence causes a stir among experts, *Science* **272**, 951 (1996).

El Naschie, M. S., Remarks on superstrings, fractal gravity, Nagasawa's diffusion and Cantorian space-time, *Chaos, Solitons and Fractals* **8(11)**, 1873-1886 (1997b).

Feigenbaum, M. J., Universal behavior in nonlinear systems, *Los Alamos Sci.*, **1**, 4-27 (1980).

Ghil, M., Cryothermodynamics: The chaotic dynamics of paleoclimate, *Physica D* **77**, 130-159 (1994).

Grossing, G., Quantum systems as order out of chaos phenomena, *Il Nuovo Cimento*, **103B**, 497-510 (1989).

Maddox, J., Licence to slang Copenhagen?, *Nature* **332**, 581 (1988).

Mary Selvam, A., Deterministic chaos, fractals and quantumlike mechanics in atmospheric flows, *Can. J. Phys.* **68**, 831-841 (1990).

Mary Selvam, A., Universal quantification for deterministic chaos in dynamical systems, *Applied Mathematical Modelling* **17**, 642-649 (1993). http://xxx.lanl.gov/html/physics/0008010.

Mary Selvam, A., J. S. Pethkar, M. K. Kulkarni and R. Vijayakumar. Signatures of a universal spectrum for atmospheric interannual variability in COADS surface pressure time series, *Int'l. J. Climatol.* **16**, 393-404 (1996).

A. Mary Selvam, Quasicrystalline pattern formation in fluid substrates and phyllotaxis. In *Symmetry in Plants*, D. Barabe and R. V. Jean (Editors), World Scientific Series in Mathematical Biology and Medicine, Volume 4., Singapore, pp.795-809 (1998). http://xxx.lanl.gov/abs/chao-dyn/9806001

Omnes, R., *The Interpretation of Quantum Mechanics*. Princeton University. Press, Princeton, NJ (1994).

Puthoff, H. E., Gravity as a zero-point fluctuation force, *Phys. Rev A* **39**, 2333 (1989).

Rae, A., *Quantum-Physics: Illusion or Reality?*, Cambridge University Press, New York (1988).

Selvam, A. M. and S. Fadnavis, Signatures of a universal spectrum for atmospheric interannual variability in some disparate climatic regimes, *Meteorology and Atmospheric Physics* **66**, 87-112 (1998a).

Selvam, A. M. and Suvarna. Fadnavis, Superstrings, Cantorian-fractal spacetime and quantum-like chaos in atmospheric flows, *Chaos, Solitons & Fractals* (in Press) (1998b)

Steinhardt, P., 1997: Crazy crystals, *New Scientist* **25 Jan**, 32-35 (1997).

Tessier, Y., S. Lovejoy, P. Hubert, D. Schertzer and S. Pecknold, Multifractal analysis and modeling of rainfall and river flows and scaling, casual transfer functions, *J. Geophys. Res.* **101(D21)**, 26427-26440 (1996).

Townsend, A. A., *The Structure of Turbulent Shear Flow*. Cambridge University Press, London, U. K. (1956).

Weinberg, S., *Dreams of a Final Theory*, Vintage, (1993).